\documentclass[a4paper]{jpconf}
\usepackage{graphicx}
\usepackage[numbers,sort&compress]{natbib}
\usepackage{amsmath}
\usepackage{amsfonts}
\usepackage{float}
\begin{document}
\title{On the comparison of optimization algorithms for the random-field Potts model}

\author{Manoj Kumar and Martin Weigel}

\address{Centre for Fluid and Complex Systems, Coventry University, CV1 5FB, England}

\ead{\\
  manojkmr8788@gmail.com\\
  Martin.Weigel@coventry.ac.uk\\
}

\begin{abstract}
  For many systems with quenched disorder the study of ground states can crucially contribute to a thorough understanding of the physics at play, be it for the critical behavior if that is governed by a zero-temperature fixed point or for uncovering properties of the ordered phase. While ground states can in principle be computed using general-purpose optimization algorithms such as simulated annealing or genetic algorithms, it is often much more efficient to use exact or approximate techniques specifically tailored to the problem at hand. For certain systems with discrete degrees of freedom such as the random-field Ising model, there are polynomial-time methods to compute exact ground states. But even as the number of states increases beyond two as in the random-field Potts model, the problem becomes {\em NP\/} hard and one cannot hope to find exact ground states for relevant system sizes. Here, we compare a number of approximate techniques for this problem and evaluate their performance.
\end{abstract}

\section{Introduction}

The presence of quenched impurities in magnetic systems can lead to fundamental shifts in material properties including topological changes in the phase diagrams such as the absence of long-range order, especially if frustration is at play \cite{young:book}. The nature of the quenched average as well as the inherently complex free-energy landscape with a multitude of metastable states separated by barriers observed for systems with strong disorder such as random-field problems and spin glasses  quickly push conventional simulation techniques to and beyond their limits \cite{janke:07}. Generalized-ensemble methods including, for instance, multicanonical simulations \cite{berg:92b}, parallel tempering \cite{hukushima:96a} and population annealing \cite{hukushima:03,machta:10a,barash:16}, do improve the situation substantially in this respect, but the run times required to equilibrate samples continue to rise rapidly with system size.

The same observation applies, in general, to the task of finding ground states of such systems, which is an {\em NP} hard problem for many discrete spin models and has similar, exponential time complexity for systems with continuous degrees of freedom. Specialist techniques borrowed from theoretical computer science can help alleviate such problems, however \cite{hartmann:book}. For the Ising spin glass, samples defined on planar graphs can be solved in a time increasing as a polynomial of the system size \cite{bieche:80a}, while the problem on non-planar and, in particular, higher-dimensional lattices is {\em NP} hard \cite{barahona:82}. The ground-state calculation for the random-field Ising model (RFIM), on the other hand, is equivalent to a minimum-cut problem which (in the absence of negative capacities that would occur for the spin glass) is in turn equivalent to the task of finding the maximum flow through an auxiliary network connecting two terminal nodes \cite{dauriac:85,hartmann:book}. Powerful methods exist for such problems \cite{ford:62,goldberg:88,kolmogorov:04a} that allow for exact calculations for large systems in any space dimension, reaching up to about $10^7$ spins in recent applications \cite{stevenson:11,fytas:13}.

These cases are benign exceptions, however, and most model extensions beyond such solvable cases quickly lead back to exponentially hard tasks. For the random-field problem, for instance, the extension from the twofold Ising to a $q$-fold Potts symmetry results in a multi-terminal flow problem that can be shown to be {\em NP} hard \cite{boykov:01}. To handle such systems, one might relax the demand of finding exact ground states and revert to stochastic approaches that provide approximate global minima with moderate effort. We have recently shown that this is possible for the case of the {\em random-field Potts model\/} (RFPM) with recourse to suitable generalizations of the {\em graph-cut} (GC) methods used for the RFIM \cite{kumar:18}. This approach relies on techniques initially developed for problems in computer vision \cite{boykov:01}. While this works quite well, there are some more recent developments in combinatorial optimization that deserve consideration as well, such as the {\it sequential tree-reweighted message passing} approach of Refs.~\cite{kolmogorov:06,kolmogorov:14}. In the present note we compare the performance of these methods for finding ground states of the RFPM.

\section{Model and methodology}
\label{model}

\begin{figure}[tb!]
  \begin{center}
    \includegraphics[width=0.9\linewidth]{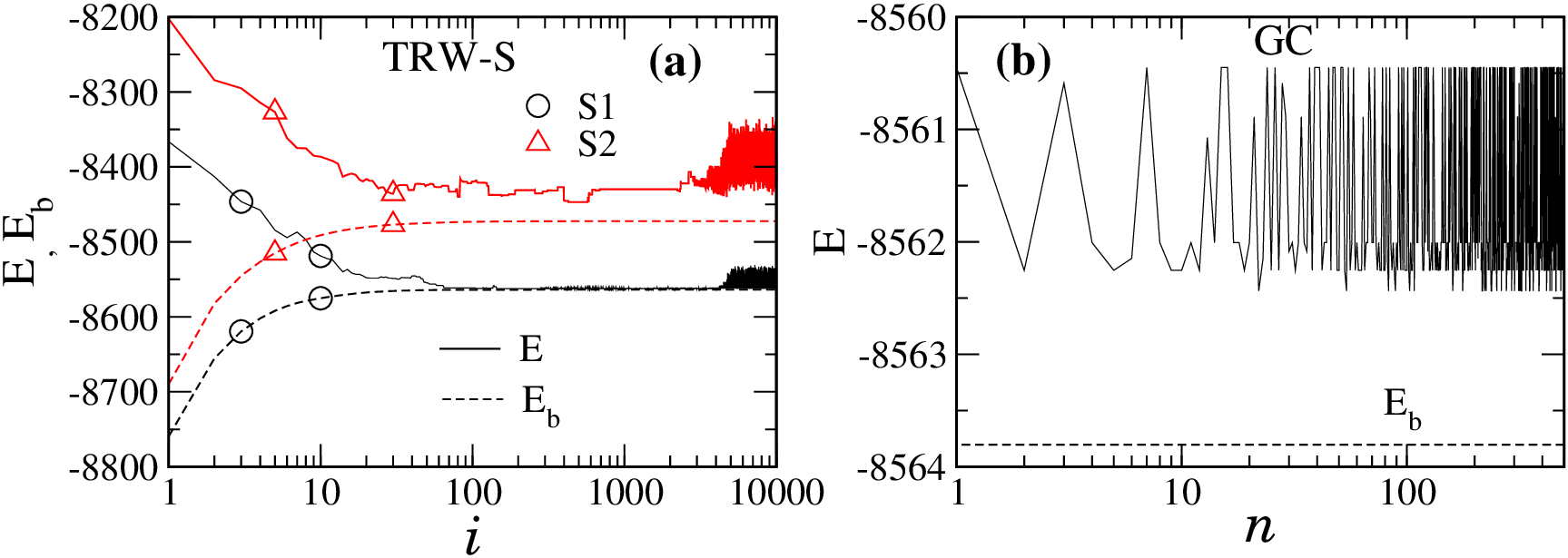}
    \caption{(a) Estimated ground-state energy $E$ and lower bound $E_b$ of two samples ($S1$ and $S2$) of the RFPM from TRW-S as a function of iteration number $i$. (b) Lowest energy $E$ of $S1$ as estimated from GC calculations using $n$ initial conditions. The dashed line corresponds to the best lower bound for $i=10\,000$ in (a). Both disorder samples are for a $64\times 64$ lattice with $q=3$ and $\Delta = 1$.}
    \label{ene_gc_trw}
  \end{center}
\end{figure}

We focus on the RFPM as defined by the following Hamiltonian~\cite{blankschtein:84},
\begin{equation}
  \label{hamilt}
  \mathcal{H}=-J\sum_{\left<ij\right>}\delta_{s_i,s_j}-\sum_i\sum_{\alpha=0}^{q-1}h_{i}^{\alpha}\delta_{s_i,\alpha},
\end{equation}
where $\delta_{x,y}$ is the Kronecker delta function. Here, the spins $s_i\in\{0,1,\ldots,q-1\}$ and $\{h_{i}^{\alpha}\}$ is the quenched random field at site $i$ and for component $\alpha$ that is drawn from a normal distribution of mean zero and standard deviation $\Delta$, i.e., $h_i^\alpha \sim {\cal N}(0,\Delta)$, such that $\Delta$ defines the strength of disorder. The problem of finding a ground state for this Hamiltonian is {\em NP\/} hard in general \cite{boykov:01} such that in practise one needs to revert to approximation techniques. Such methods have been considered quite extensively in computer vision, where optimization problems very similar to the RFPM ground-state problem occur in the context of image restoration, motion and stereo \cite{boykov:01,kolmogorov:04a,kolmogorov:06}.

One of the most popular approaches was proposed by Boykov {\em et al.\/}~\cite{boykov:01} based on the application of the graph-cut method to embedded binary decision problems. They considered two variant algorithms dubbed $\alpha$-$\beta$-swap and $\alpha$-expansion, respectively, that are applied to images composed of pixels carrying one of $q$ labels (such as colors, for example). For the $\alpha$-$\beta$-swap, two labels $\alpha \ne \beta \in \{0,1,....,q-1\}$ are picked and all other labels are frozen; the update consists of a swap of the labels between  regions. In contrast, for $\alpha$-expansion one picks a label $\alpha$ and attempts to expand it while freezing all the remaining ones, cycling through the labels in turn in $q$ iterations. In both cases, the resulting embedded binary (Ising type) problems are solved exactly using the established techniques for maximum-flow \cite{ford:62,goldberg:88,kolmogorov:04a}. For our application to the RFPM we focus on the $\alpha$-expansion move which we have shown previously to be quite efficient for finding approximate ground states of this system \cite{kumar:18}. By construction, this approach is not guaranteed to result in a ground state, but it will normally lead to a metastable configuration. The approach is stochastic in that the final state depends on the chosen initial spin configuration. As a consequence, results can be systematically improved by performing $n$ independent runs with random initial conditions and picking the result of lowest energy \cite{weigel:06b,kumar:18}. Since $\alpha$-expansion cannot increase the energy, this approach is guaranteed to result in the exact answer if (but not only if) the ground state was among the initial configurations. While this appears unlikely for non-trivial system sizes, it indicates that such a procedure should converge in the limit $n\to\infty$.

\begin{figure}
  \begin{center}
    \includegraphics[width=0.95\linewidth]{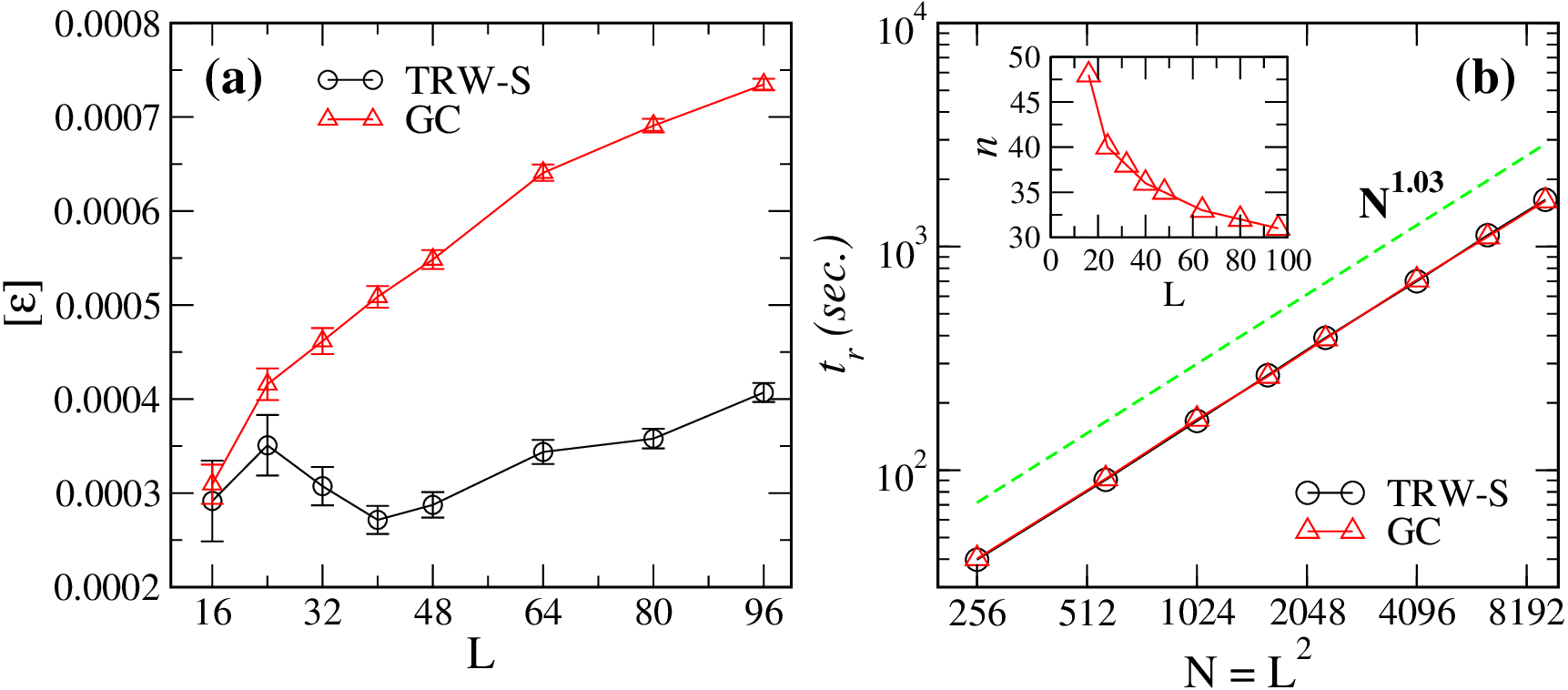}
    \caption{%
      (a) Accuracy $[\varepsilon]$ according to Eq.~\eqref{acc} averaged over 1000 disorder samples against system size $L$ after fixing $i=1000$ for TRW-S and varying $n$ of GC to match the run time $t_r$ of TRW-S plotted in panel (b). The dashed line corresponds to the power-law increase $\sim N^{1.03}$ resulting from a fit to the data. The inset shows the number of initial conditions used for GC to lead to run time $t_r$.}
    \label{acc_long_run}
  \end{center}
\end{figure}

A different class of approximate methods was considered by Kolmogorov in Ref.~\cite{kolmogorov:06}. It is based on ideas relating to belief propagation and the more general concept of message passing  \cite{mezard:09}, where ``messages'' with weights updated according to the local marginal distributions are passed along edges of a graph to iterate probabilistic graphical models until convergence is achieved. Such techniques are exact for trees, but do not guarantee global minima in the presence of loops. The method of (sequential) {\em tree-reweighted message passing\/} (TRW-S) proposed in Ref.~\cite{kolmogorov:06} uses a specific way of covering the graph with trees to approximate the solution. As a byproduct, based on a relaxation for a dual linear programming problem, it also maintains a (rigorous) lower bound $E_b$ to the value of the cost function. TRW-S improves with the number of iterations employed, but in contrast to GC it is not known how to systematically improve it by using repeated runs \cite{kolmogorov:private}. In most cases, the proposed solution has higher energy than the lower bound, but it is clear that if the two energies coincide an exact ground state has been found. In practise, we run TRW-S for a large number of iterations $i$, and pick the state of lowest energy encountered.

\section{Numerical results}
\label{result}

We studied the two-dimensional RFPM for $q=3$ on a periodic square lattice of edge length $16 \le L \le 96$ in order to assess and compare the performance of the GC and TRW-S algorithms discussed above. In the following we focus on $\Delta=1$, corresponding to quite strong disorder for this 2D system \cite{kumar:18}. As demonstrated in Ref.~\cite{kumar:18}, such sizes  are large enough for a non-trivial benchmark. For a given disorder sample $\{h_i^\alpha\}$, we vary the number of iterations $i$ of TRW-S (up to a maximum of 10\,000), and  the number $n$ of random initial spin configurations for GC (up to a maximum of 1000). This is illustrated in Fig.~\ref{ene_gc_trw} for two samples of size $L=64$. In TRW-S (panel a) while the energy of the current configuration fluctuates, the value of the lower bound increases monotonically, consistent with Ref.~\cite{kolmogorov:06}. For GC, on the other hand, the value of the lowest energy found in $n$ independent runs fluctuates and decreases only slowly with $n$ (panel b).

\begin{figure}
  \begin{center}
    \includegraphics[width=0.95\linewidth]{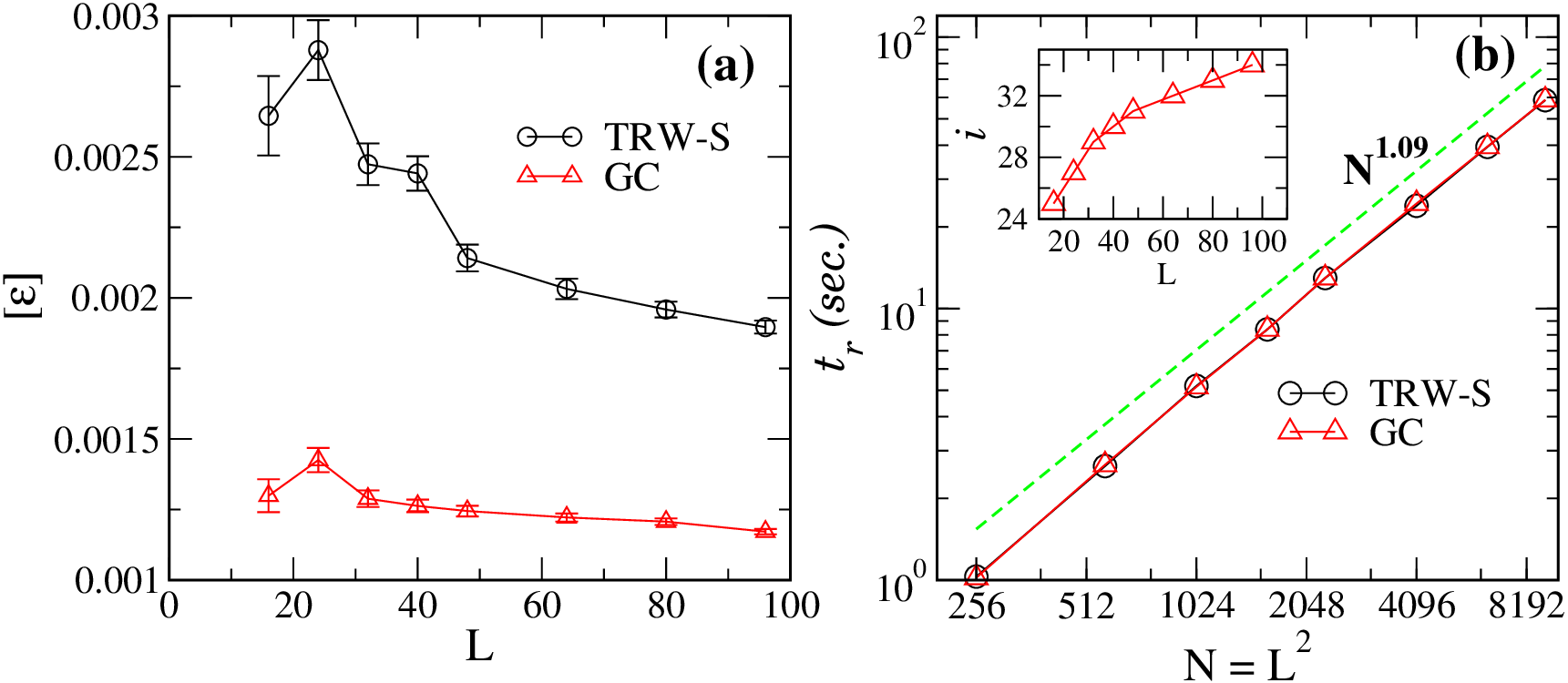}
    \caption{(a) Disorder averaged accuracy $[\varepsilon]$ as a function of $L$ while fixing $n=1$ of GC and varying $i$ of TRW-S. (b) Run times for this setup. The inset shows the number of TRW-S iterations required. The green dashed line corresponds to the run-time scaling of GC, $t_r\sim N^{1.09}$.}
    \label{acc_short_run}
  \end{center}
\end{figure}

To compare the two approaches more systematically, we tuned both algorithms to result in equal run times, which can be achieved by fixing the number of iterations $i$ of TRW-S or varying the number of initial conditions $n$  of GC, respectively. Performance is assessed by considering the accuracy parameter
\begin{equation}
 \varepsilon=\frac{E_b-E_{\min}}{E_b},
 \label{acc}
\end{equation}
where $E_{\min}$ is the best estimate of the ground-state energy from TRW-S or GC, and $E_b$ is the largest value of the lower bound to the ground-state energy of a sample found by TRW-S (which is achieved at $i=10\,000$). In Fig.~\ref{acc_long_run}(a) we compare the disorder averaged accuracy $[\varepsilon]$ between GC and TRW-S for a range of system sizes. Here, the GC method is tuned in $n$ to match the run time $t_r$ of TRW-S for $i=1000$ as shown in panel (b) and its inset. As it is seen there, the total run time is nearly linear in the system volume, a power-law fit resulting in an estimate $t_r \sim N^{1.03}$. The inset shows how the number of initial conditions $n$ required to lead to identical run time with TRW-S depends on $L$. We also considered a complementary setup, where $n=1$ was fixed for GC and the number $i$ of iterations of TRW-S was tuned to result in the same run time. The result is shown in Fig.~\ref{acc_short_run}. The behavior of run times shown in panel (b) is clearly consistent with a linear scaling in $N$, in line with previous findings in Ref.~\cite{kumar:18}. From the comparison of $[\varepsilon]$ shown in panel (a) we see that, interestingly,  GC outperforms TRW-S at $n=1$, meaning that the GC technique finds a quick approximate solution.

\begin{figure}
  \begin{center}
    \includegraphics[width=0.95\linewidth]{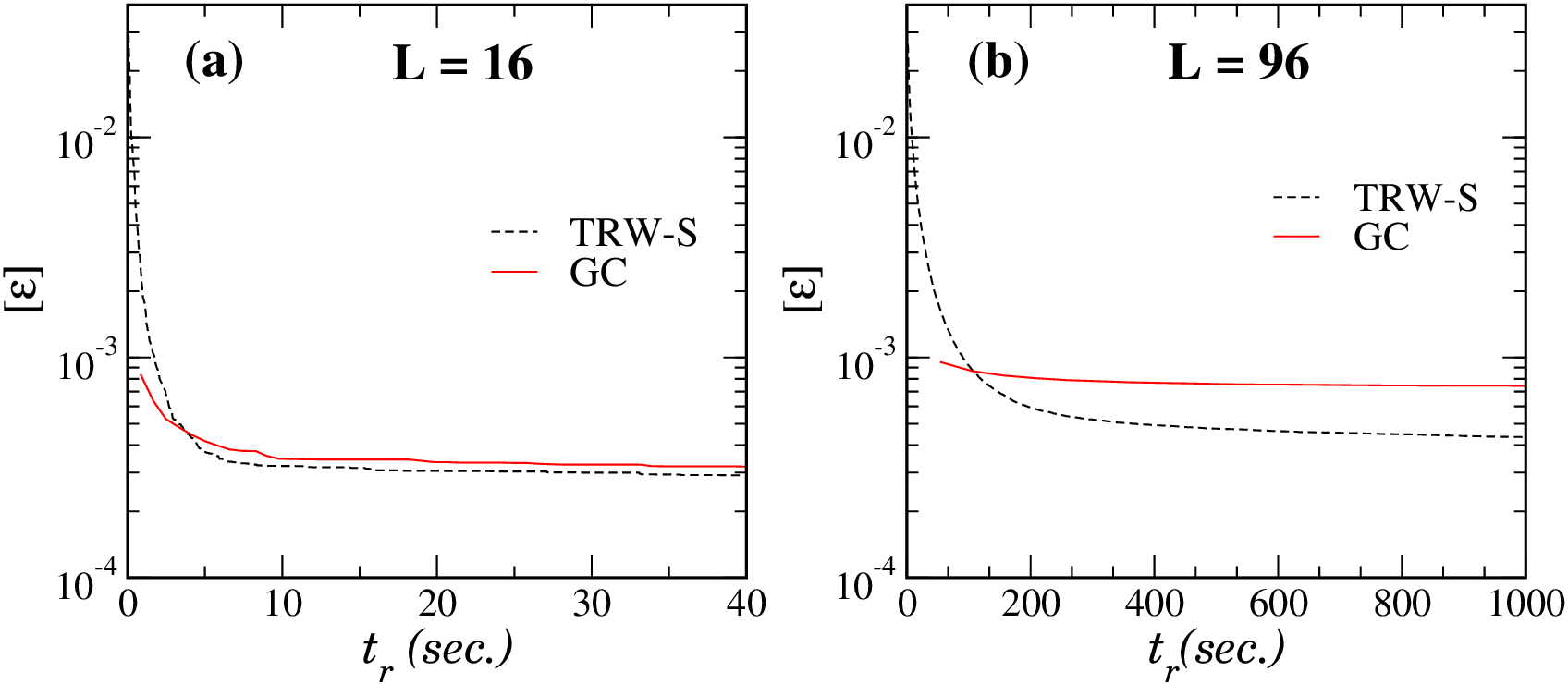}
    \caption{Accuracy $[\varepsilon]$  averaged over 1000 samples as a function of run time $t_r$ for TRW-S and GC. Panel (a) is for $L=16$ whereas panel (b) is for $L=96$.}
    \label{acc_run_time}
  \end{center}
\end{figure}

Finally, it is instructive to consider the quality of approximation as a direct function of the run time invested. In Fig.~\ref{acc_run_time}, we show the comparison of $[\varepsilon]$ for a fixed $L$ but with increasing $t_r$. It is apparent that initially GC finds better solutions than TRW-S, but with increasing time investment there is a crossover and eventually TRW-S outperforms GC.

\section{Summary and conclusions}
\label{conclude}

We studied the performance of the TRW-S and $\alpha$-expansion GC methods for the two-dimensional random-field Potts model. Both techniques have been tuned to match their run times. Such tuning can be done in two possible ways, i.e., either by varying the number of initial conditions $n$ of GC to match the run time $t_r$ for $i$ iterations of TRW-S, or vice versa. For a fixed number of iterations for TRW-S resp.\ a fixed number of initial conditions for GC, the run time increases linearly with the number of spins. Investigating both types of comparisons, we find that GC is quicker in finding a reasonable approximation in a short time, but TRW-S leads to better overall results when more significant time is invested.  These findings are consistent with the study of Kolmogorov \cite{kolmogorov:06}, who compared these techniques for a stereo matching problem. While hence TRW-S could be considered the preferable technique, an important advantage of GC is that it converges for $n\to\infty$, albeit slowly \cite{kumar:prep}. A similar extrapolation parameter is not known for TRW-S \cite{kolmogorov:private}, suggesting an interesting avenue for future research.

\ack The authors acknowledge support by the Royal Society--SERB Newton International
Fellowship (NIF$\backslash$R1$\backslash$180386). We acknowledge the provision of
computing time on the parallel compute cluster Zeus of Coventry University.

\bibliography{JPCS_rfpm_new_ver}

\end{document}